\documentstyle[prl,aps,epsfig,twocolumn,floats]{revtex}
\begin{document}
\draft

\title{\hfill OKHEP-00-04\\Improved Experimental Limits on the Production of  
Magnetic Monopoles}
\author{G. R. Kalbfleisch,\thanks{Electronic mail: {\tt grk@mail.nhn.ou.edu}}
 K. A. Milton, M. G. Strauss, L. Gamberg, 
E. H. Smith,\thanks{Present address: Lockheed Martin,
Palo Alto, CA 94304} and W. Luo}
\address{Department of Physics and Astronomy,
University of Oklahoma,
Norman, Oklahoma 73019}
\date{\today}
\maketitle
\begin{abstract}
We present new limits on low mass accelerator-produced point-like Dirac
magnetic monopoles trapped and bound in matter surrounding the D\O\ collision 
region of the Tevatron at Fermilab (experiment E-882).  In the
context of a Drell-Yan mechanism,
we obtain cross section limits for the production of monopoles with
magnetic charge values of 1, 2, 3, and 6 times the 
minimum Dirac charge of the order of picobarns, some hundred times smaller 
than found in similar previous Fermilab searches.
Mass limits inferred from these cross section limits are presented.
\end{abstract}

The existence of magnetic monopoles of charge $g$ explains the quantization 
of electric charge $e$ as  $eg=n \hbar c/2$, $n=\pm1$, $\pm 2$, \dots \cite{1},
results in the dual-symmetrization of Maxwell's equations \cite{2},
and is not forbidden by any known principles of physics. 
The minimum values of the quantization number are $n=1$ according to Dirac and
$n=2$ according to Schwinger \cite{1}. For $e$ being the charge on the
electron, these values are $n=3$ and $n=6$,
respectively, if quantization with the quark electric charges is allowed.
 Previous searches  
for trapped and bound magnetic monopoles in samples from various accelerators 
\cite{3,4}, in meteorites \cite{5}, and lunar soil 
\cite{7} have been made.  Since the Tevatron 
proton-antiproton ($p\bar p$) collider has extended its integrated 
luminosity by a factor of some ten thousand over the last search of 
Bertani et al.~\cite{4}, we have taken samples exposed in the D\O\ experiment 
and performed a search that improves these limits.  We use 
the induction method of Alvarez et al.\ \cite{8} to measure the magnetic 
charge content of macroscopic Al and Be samples.  
Indirect searches \cite{9} are able to present limits beyond those 
energetically 
allowed in direct searches, but represent an entirely different set 
of assumptions as to their observation, and are not further discussed in 
this letter.  We also do not discuss very 
massive GUT monopoles, such as those searched for in cosmic rays \cite{10}.

This extension of limits is experimentally driven.  Theoretical motivation, 
beyond the general principles given above, derives from the possibility that 
magnetic monopoles generated during spontaneous symmetry breaking
at the electroweak scale might give rise to monopoles of mass $\sim2.5$ 
to $\sim15$ TeV \cite{11}.  The search here can only raise the 
previous limits of $\sim0.1$ TeV to $\sim0.4$ TeV.  At the Large Hadron 
Collider one could approach 2 TeV with the same techniques.

A large warm bore cryogenic detector, similar to that of Longo
et al.\ \cite{5}, was constructed and operated at the University of Oklahoma.
The active elements of the detector are two
19 cm diameter superconducting loops each connected to a
DC SQUID (Superconducting QUantum Interference Device) \cite{12}.
The Meisner effect prevents a change in the net 
flux through the loops, resulting in a change, or ``step,''
in current flowing in the 
loops whenever a magnetic charge passes through them. Samples of a 
size less than 7.5 cm in diameter by 7.5 cm in length are repeatedly passed
through the 10 cm diameter warm bore centered on and perpendicular
to the loops, traveling some 108 cm about the position of the 
superconducting loops.  In a central 64 cm region this allows
for the magnetic effects of induced 
and permanent dipole moments in the sample  to start 
and return to zero on each up and each down traversal of the sample.  
High frequency noise is filtered out, while
random and $1/f$ noise contributions are averaged out over 20 
pairs of up/down passes.  Each up or down traversal takes some 25 
seconds.  A net data rate of 10 Hz is 
recorded for each of the two SQUIDs; recorded as well are the readings from an 
accelerometer, the vertical position from an optical encoder, 
the number of increments taken by the stepper motor, and the time.  

The SQUIDs are tuned and their transfer functions measured periodically 
according to the manufacturer's specifications in order to keep them 
operating with constant sensitivity. 
The absolute calibration of an expected signal from a Dirac monopole was made 
using a ``pseudopole.''  A long thin magnetic solenoid (1.5 cm diameter 
by 100 cm length with 4710 turns per meter) carrying a calibrated small 
current gives a calculable pseudopole at either end.  This pseudopole
can either be passed through the warm bore of the detector in 
a way similar to the 
samples, or it can be placed in a given position with one end fully extended 
through the SQUID loops and the solenoid current repeatedly
switched on and off.
Both methods were used and a value of $2.40 \pm 0.04$ mV/minimum Dirac charge 
was obtained.  The internal random and systematic local noise 
contributions were typically 0.2 mV, although the external 
systematic error of $\sim0.7$ mV dominates.

The shape of the pseudopole curve is compared to that of a theoretical 
calculation in Fig.~1a and the response to a point magnetic dipole  is
compared to a narrow one from the data.    Fig.~1b
shows the step from a 5.5mV (i.e., 2.3 Dirac 
poles) signal from a run of a pseudopole embedded in an Al sample.
The movement of conducting samples through the
small residual magnetic field gradient in the warm 
bore also gives rise to an induced dipole, which cancels out over an up/down 
pair.

\begin{figure}[ht]
\centerline{\psfig{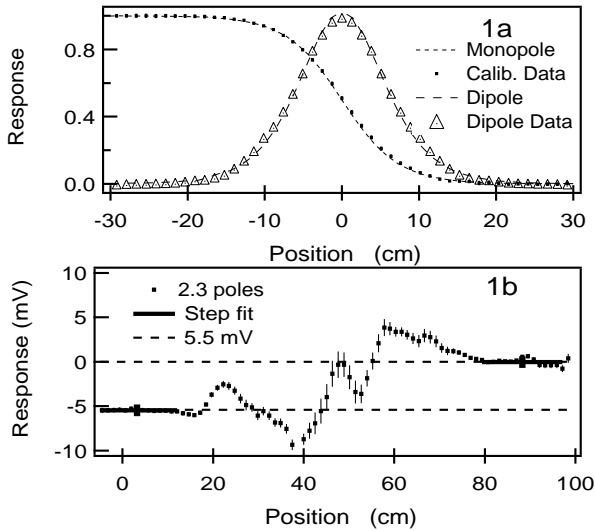}}
\caption{``Pseudopole'' curves.  
a) Comparison of theoretical monopole response  to an 
experimental calibration and of
 a simple point dipole of one sample with that calculated 
from the theoretical response curve.  b) The observed 
``step'' for a pseudopole current, corresponding to 2.3 minimum 
Dirac poles, embedded in an Al sample.}
\label{fig1}
\end{figure}

There are two kinds of samples, beryllium and aluminum.  A 46 cm 
section of the 5 cm diameter D\O\ Be beam pipe, centered on the collision 
region, covering nearly the full solid angle, was cut into six 7.6 cm pieces.  
Two extension cylinders of Al, each of 150 cm diameter by 46 cm length and of 
1.26 cm thickness, 
also were cut into pieces of 7 cm by 7.6 cm or of 6 cm by 7.6 
cm, and bundled four pieces to a ``sample.''  
The effective solid angle acceptance \cite{15} subtended by both extension 
cylinders was 0.12 of $4 \pi$, around a cosine of the laboratory polar angle of 
$-0.82$ or 0.82.  There were a total of 222 Al samples.  Both the Be and 
Al samples were measured using the aforementioned traversal scheme.  Since a 
nylon string, which could be magnetically contaminated, pulled the 
samples through the warm bore, we also ran background runs
between every two samples. 

The data analysis proceeded as follows.  The time sequences of the SQUIDs'
outputs 
were examined interactively, and bad sections deleted pairwise (up/down); 
typically 17--18 pairs of traversals remained.  A pedestal value 
(the SQUID output near the top end of each traversal) was 
subtracted from every voltage value along that up (or down) traversal.  
The values for each of some 90 small 
ranges of vertical positions were averaged,
removing most of the random drift of the SQUIDs.  
The two SQUIDs' data were averaged, shifting one relative to the other by
10.1 cm in position in order to superimpose their dipole responses.
The background samples 
were analyzed similarly and local groups of background runs were 
averaged.  These backgrounds were subtracted from the samples'
spectra.  A horizontal line was fit to two regions, one at the lower 
position and one at the upper, as seen earlier in Fig.~1b; the difference 
in the values of these two flat fits gives the step
for that sample.   Examples of sample spectra are shown in Fig.~2a and 2b.  
The steps from the various samples are histogramed in Fig.~3.  The 
background subtraction ensures that the distribution of steps centers on zero, 
since the small effect of the magnetized string holding the 
sample has been removed. 

\begin{figure}[ht]
\centerline{\psfig{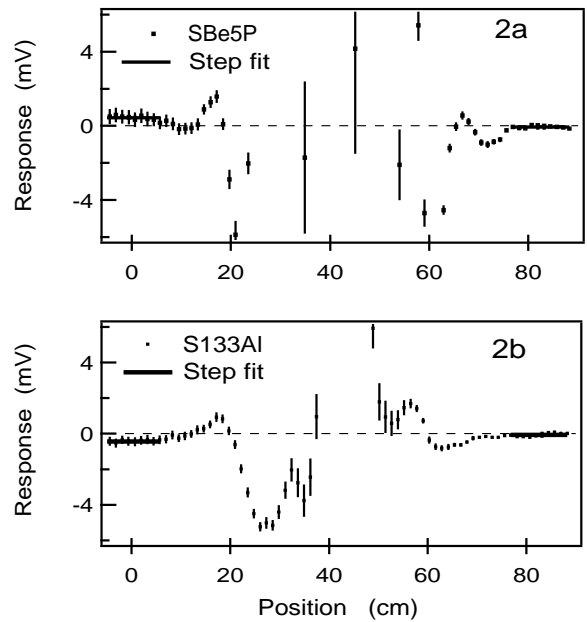}}
\caption{Sample spectra. a)  
Beryllium sample ``SBe5P,'' and b) aluminum 
sample ``S133Al.'' The observed steps are $-0.8$ mV in a) and 
$+0.4$ mV in b).  The dipole signals are off scale in the middle
regions of the plot in this vertically expanded view.}
\label{fig2}
\end{figure}

The distribution of steps for the data has a mean of 0.16 mV and
an rms spread (sigma) of 0.73 mV, as shown in Fig.~3.
One sided 90\% confidence limits for monopole charges of $n=1$ or
$-1$ can be obtained by considering the number of samples that are within
1.28 sigma of the $n=\pm1$ positions, corresponding to samples
outside of the central region of $\pm1.47$ mV around zero.  
We find 8 samples outside this
central region, where 10.4 were to be expected from the Gaussian error.  
According to Feldman and Cousins \cite{16}, 
the 90\% confidence upper limit is 4.2 signal events
 for 8 events observed when 10 were expected. 
In order to be sure that none of the 8  outlying  samples were 
monopole candidates, 
we remeasured them along with a few control samples; all eight 
remeasured samples fell within $\pm1.47$ mV of $n=0$. 

No samples are within 1.28 sigma of the $|n|\ge2$ positions, the 
closest being 3.08 sigma away from $n=-2$. The 90\% confidence upper limit 
is 2.4 signal events for zero events observed and zero expected.
The upper limit numbers 4.2 and 2.4  are used to 
derive cross section upper limits for $|n|=1$ and $|n|\ge2$, respectively.

\begin{figure}[ht]
\centerline{\psfig{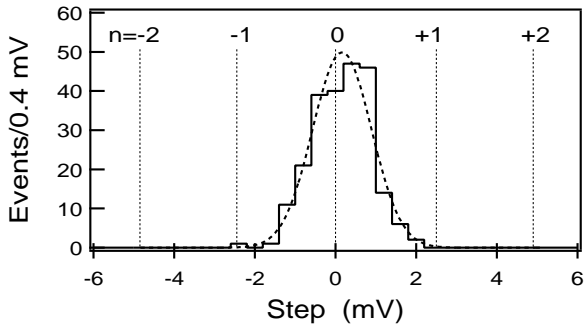}}
\caption{Histogram of steps.  
Vertical lines (dashed) define the expected positions of signals for 
various $n$.  The Gaussian curve 
(dashed) corresponds to 228 measurements having an average 
value of 0.16 mV and an rms sigma of 0.73~mV.}
\label{fig3}
\end{figure}

The acceptance of the experiment to the stopping and trapping of monopoles 
must be estimated.  It is a function of ranging out of monopoles 
due to the energy loss
in bulk matter and the distribution in energy of the produced 
monopoles.  It is assumed that ranged out and stopped monopoles bind to the 
magnetic dipole moments of the appropriate nuclei, $^9$Be or 
$^{27}$Al \cite{17}.
The production is assumed to derive from a Drell-Yan process: quark-antiquark
annihilation to monopole-antimonopole pair via an intermediate high mass 
virtual photon.  The shape of the energy distribution follows  from a 
dimensional argument that is basic to Drell-Yan: $M^3d\sigma/dM$ 
is dimensionless, where $M$ is the invariant mass 
of the pair of monopoles. This $p\bar p$ cross section must include a
threshold phase space factor and the velocity dependence of the monopole 
interaction.  The threshold factor is $\beta$, the velocity of the 
monopole in the CM system. We take the interaction factor to be
$\beta^2$, since the Lorentz force for magnetic charges $g$ is  
${\bf F} = g ({\bf H}  -  {\bf v \times D})$.  
Thus the energy shape of $d\sigma/dM$ is $(\beta/M)^3$,
convoluted with the momentum distributions of the quarks in the colliding
proton and antiproton.
In the absence of a theory of monopole cross sections, we are led to make
these model assumptions. Fig.~4a 
shows the shape for the cases $\beta=1$ and $\beta=v/c$.  The acceptance is the 
area of this distribution between two $M$ limits divided by the area of the 
total distribution.  The two $M$ limits are  $M_{\rm low}= 2m + 2T_{\rm low}$
and $M_{\rm high}= 2m + 2T_{\rm high}$, where $m$ is the mass of each 
produced monopole and the $T$'s are the kinetic energies of monopoles which 
are either just entering or just ready to exit the 
sample.  If the energy is too low or too high, the monopole
will not be absorbed by that part of the detector.  The energy loss 
functions as summarized by Ahlen \cite{18} 
were parameterized as to range, monopole mass, angle, and sample.  The
estimated  acceptances given in Table I vary 
slowly in the mass regions of interest.

Using the total luminosity delivered to D\O, $172\pm 8$ 
pb$^{-1} \cite{18.5}$, the number limit of monopoles, the acceptance, and the 
solid angle coverage given above, we obtain the $p\bar p$ cross section limits
shown in Table I.  These limits are more than 100 times smaller than the best 
prior Tevatron limit of Bertani et al.\ \cite{4} of 200 pb.

\begin{figure}[ht]
\centerline{\psfig{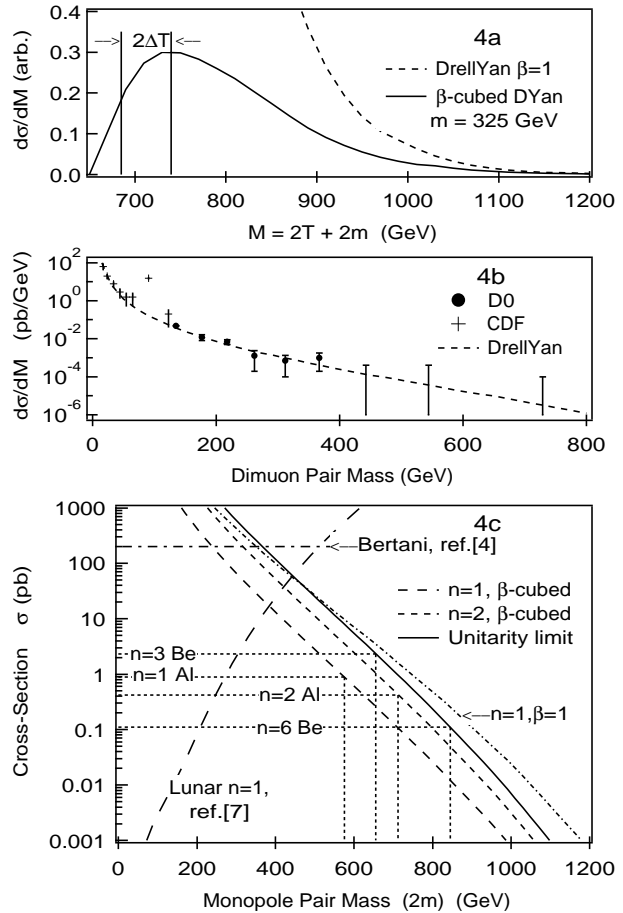}}
\caption{Cross-section plots. a)  $d\sigma/dM$ for 
Drell-Yan production.  The solid curve, cut between the two kinetic energy ($T$) limits, 
yields the acceptance for stopping and trapping $n=1$
monopoles in the aluminum 
extension cylinder.  The dashed curve gives that for dimuons. 
b) The differential dimuon cross section is normalized to the 
D\O\ \protect\cite{20} and CDF \protect\cite{19} dimuon data.
c) The $d\sigma/dM$ curve of b) is multiplied by $\beta^3$, 
integrated and renormalized by a factor of $n^2(137/2)^2$ to give the 
cross section curves versus pair mass ($2m$) as shown.  
 For the lunar samples, the ordinate is the proton-nucleon cross section.
  The $n=1$, $\beta=1$ curve is shown for
comparison because the $\beta^3$ correction
was not included in previously published limits.}
\label{fig4}
\end{figure}

One can further interpret these limits as mass limits using
the scaled Drell-Yan cross sections.  Here the 
cross section is taken to be  $n^2 (137/2)^2$ larger than the Drell-Yan 
muon pair cross section, modified by $\beta^3$,
for $p\bar p$ interactions measured by 
CDF \cite{19}  and by D\O\ \cite{20}. For such large
cross sections a unitarity limit appears
at an equivalent $ n^2 \sim 9$ \cite{21}.  We thus use the $n=1$ or 2 
scalings for the cases $n=1$ or 2, and the unitarity limit for higher $n$ 
values in converting cross section limits into mass limits.  We offer 
this procedure, as well as that for determining the cross sections,
as canonical. We realize that future theoretical 
work may change this interpretation of the data.  The 
normalization of the Drell-Yan muon pair calculation (using the CTEQ5m parton 
distribution functions \cite{22}) to the CDF and D\O\ data are shown in Fig.~4b.  
The result of the integration of $d\sigma/dM$ to a total cross section for 
monopoles is shown in Fig.~4c, along with the upper limits for the data of 
Table I.  The Bertani et al.\ \cite{4} and the lunar rock sample limits
of Ross et al.\  \cite{7} are also shown. The corresponding interpreted
lower mass limits are given in Table I.  These limits are some  3 times 
larger than those of prior searches for accelerator-produced
monopoles trapped in matter.

\section*{Acknowledgements}
We acknowledge the support of the US Department of Energy.  We thank OU's
Departments of Physics and Astronomy and of Aeronautical and 
Mechanical Engineering for support,
S. Murphy, J. Young and the Physics Machine Shop, and
A. Wade of the University of Oklahoma, M. Longo of the University of Michigan, 
and T. Nicols, M. Kuchnir, 
and H. Haggerty of Fermilab. 
We thank Fermilab and the D\O\ collaboration for 
the samples.  REU students I. Hall and W. Bullington participated in this
work during the summers.

\newpage
\widetext

\begin{table}
\begin{tabular}{ccccc}
Magnetic Charge&$|n|=1$&$|n|=2$&$|n|=3$&$|n|=6$\\
\hline
Sample&	Al&Al&Be&Be\\
$\Delta\Omega/4\pi$ acceptance&0.12&0.12&0.95&0.95\\
Mass Acceptance&0.23&0.28&0.0065&0.13\\
Number of Poles&$<4.2$&$<2.4$&$<2.4$&$<2.4$\\
Upper limit on cross section&0.88 pb&0.42 pb&2.3 pb&0.11 pb\\
Monopole Mass Limit&$>285$ GeV&$>355$ GeV&$>325$ GeV&$>420$ GeV
\end{tabular}

\caption{Acceptances, upper cross section limits, and lower mass limits,
as determined in this work (at 90\% CL).}

\end{table}

\begin{references}
\bibitem{15} We take into account the longitudinal momentum of the 
quark-antiquark pair, and assume the production is isotropic in the
monopole-pair center of mass frame.
\bibitem{21}  We equate the elementary quark-antiquark reaction cross section 
to $(2J+1)\pi/s$ for $J\sim1$.
\end{references}
\end{document}